# Beam dynamics corrections to the Run-1 measurement of the muon anomaly $a_\mu$ at Fermilab

**Alessandra Lucà**[a,*] **on behalf of the Muon $g-2$ Collaboration**

[a]*Fermi National Accelerator Laboratory, Batavia, IL, USA*

  E-mail: aluca@fnal.gov

The Fermi National Accelerator Laboratory (FNAL) Muon $g-2$ Experiment has measured the positive muon magnetic anomaly $a_\mu \equiv (g_\mu - 2)/2$ with a precision of 0.46 parts per million, with data collected during its first physics run in 2018. The experimental result combined with the measurement from the former experiment at Brookhaven National Laboratory increases the tension with the Standard Model expectation to $4.2\sigma$, thus strengthening evidence for new physics. The magnetic anomaly is determined from the precision measurements of the muon spin precession frequency, relative to the muon momentum, and the average magnetic field seen by the beam. In an ideal case with muons orbiting in a horizontal plane with a uniform vertical magnetic field, the anomalous precession frequency $\omega_a$ is given by the difference between the spin ($s$) and cyclotron ($c$) frequencies, $\omega_a = \omega_s - \omega_c$. The observable $\omega_a$ is proportional to $a_\mu$.

This proceeding presents the beam dynamics systematic corrections that are required to adjust the measured muon precession frequency $\omega_a^m$ to its true physical value $\omega_a$.

[*]Speaker

## 1. Introduction

The Fermilab Muon $g-2$ Experiment (E989) reports a new measurement of the positive muon magnetic anomaly $a_\mu \equiv (g_\mu - 2)/2$, where $g_\mu$ is the gyromagnetic ratio of the muon [1]. The Run-1 result is $a_\mu = 116\,592\,040\,(54) \times 10^{-11}$ (0.46 ppm), where the total uncertainty includes the dominant statistical uncertainty combined with the precession-rate systematic [2], magnetic systematic [3], and beam-dynamics systematic uncertainties [4].

This experiment measures $a_\mu$ by observing the anomalous spin precession frequency $\omega_a$ of a muon ensemble within a storage ring. In an ideal case with muons orbiting in a horizontal plane within a uniform vertical magnetic field, the anomalous precession frequency $\omega_a$ is given by the difference between the spin ($s$) and cyclotron ($c$) frequencies, $\omega_a = \omega_s - \omega_c$. Eq. 1 gives the rate of change of the component of spin $\vec{S}$ parallel to the velocity $\vec{\beta} = \vec{v}/c$ in the presence of an electric field $\vec{E}$ and a magnetic field $\vec{B}$.

$$\frac{d(\hat{\beta} \cdot \vec{S})}{dt} = -\frac{q}{m}\vec{S}_T \cdot \left[ a_\mu \hat{\beta} \times \vec{B} + \beta \left( a_\mu - \frac{1}{\gamma^2 - 1} \right) \frac{\vec{E}}{c} \right], \tag{1}$$

where $q$, $m$ and $\gamma$ are the muon charge, mass, and Lorentz factor respectively, and $\vec{S}_T$ is the component of $\vec{S}$ perpendicular to $\hat{\beta}$. The electric field term in Eq. 1 vanishes for muons having the "magic" momentum $p_0 = 3.094\,\mathrm{GeV}/c$ ($\gamma \sim 29.3$). Thus, the experiment is designed around injection and storage of muons centered on $p_0$.

The Fermilab Muon Campus [5] delivers sixteen highly polarized positive muon ($\mu^+$) beam bunches every 1.4 s into a 14.2-m-diameter superconducting storage ring (SR), which has a highly uniform $\sim 1.45\,\mathrm{T}$ vertical magnetic field. After traversing one quarter of the SR, the beam is deflected by a fast pulsed-kicker magnet system [6] onto the intended storage orbit. Four electrostatic quadrupole (ESQ) sections [7] installed symmetrically around the ring confine the beam vertically. The Run-1 data, collected in 2018, are grouped into four datasets (a-d) corresponding to four different kicker magnet and ESQ voltage combinations. The $\omega_a$ frequency is encoded in the modulation of the decay-positron energy spectrum as muons circulate in the ring. Because of parity violation in the $\mu^+$ weak decay, positrons are emitted with an energy and an angular distribution that are each correlated to the muon spin direction in its rest frame. Ignoring effects from beam dynamics, 24 calorimeter stations [8] distributed uniformly around the inner radius of the SR see a positron count rate versus time-in-fill $t$ and positron energy $E$ with the following equation:

$$N(t, E) = N_0(E) e^{-t/\gamma\tau_\mu} \left\{ 1 + A(E) \cos[\omega_a t + \varphi_0(E)] \right\}. \tag{2}$$

The normalization, time-dilated muon lifetime, asymmetry, anomalous precession frequency, and ensemble average phase at the time of injection are represented in Eq. 2 by $N_0(E)$, $\gamma\tau_\mu$, $A(E)$, $\omega_a$ and $\varphi_0(E)$, respectively. The frequency extracted by fitting the data is the measured quantity $\omega_a^m$, which must be corrected for perturbations from beam dynamics effects. Two straw-tracker stations [9] are critical to the beam dynamics study since they provide detailed time-dependent stored-muon spatial profiles in two areas of the SR. The beam dynamics corrections are identified as electric field $C_e$, pitch $C_p$, muon loss $C_{ml}$, and phase acceptance $C_{pa}$. They are applied in a linear combination as $\omega_a \approx \omega_a^m \cdot (1 + C_e + C_p + C_{ml} + C_{pa})$ and are further discussed in Sec. 2 and in Sec. 3.

## 2. Electric Field Correction $C_e$ and Pitch Correction $C_p$

The electric field term in Eq. 1 produces a rest frame magnetic field that affects the measured anomalous precession frequency $\omega_a^m$. For the simple case, where we neglect the vertical betatron motion, Eq. 1 leads to

$$\omega_a^m = \frac{|q|}{m} a_\mu B_y \left[ 1 - \beta \frac{E_r}{cB_y} \left( 1 - \frac{m^2 c^2}{a_\mu p^2} \right) \right], \qquad (3)$$

where the subscripts $r$ and $y$ denote the radial and vertical components, respectively[1]. The electric field term vanishes at the magic momentum, or when $E_r = 0$, which is the case of the central orbit radius $R_0$ as a result of the design of the ESQ system. In practice, the stored muon distribution has a finite momentum spread and is not well-centered at $R_0$. A fast rotation analysis using the Fourier method yields the distribution of cyclotron frequencies $f$, which we convert to equilibrium radii $R$ through the relation $R(2\pi f) = v$. We assume a fixed muon velocity $v$. Fig. 1a shows the distributions of equilibrium radial offsets $x_e := R - R_0$. The electric field correction depends on the mean and width of these distributions via $C_e \approx 2n(1-n)\beta_0^2 \frac{\langle x_e^2 \rangle}{R_0^2}$, where $\langle x_e^2 \rangle = \sigma_{x_e}^2 + \langle x_e \rangle^2$. The field index $n$ is defined by $n = \frac{R_0}{vB_0} \frac{\partial E_y}{\partial y}$, with the field gradient determined by the quadrupole voltages and geometry. Finally, the electric field correction is $C_e = 489 \pm 53$ ppb.

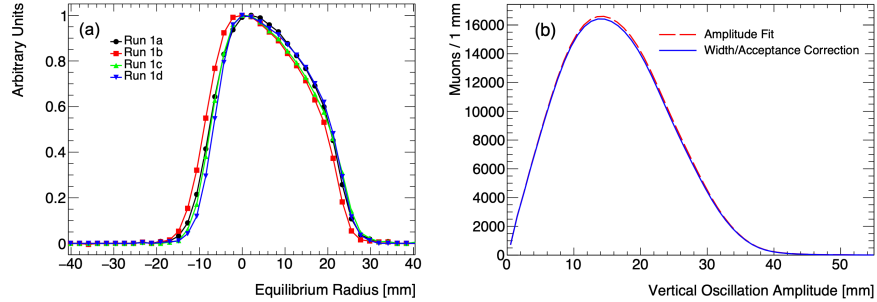

**Figure 1:** (a) The radial distribution extracted by the Fourier method for the four Run-1a datasets. The equilibrium radius is defined such that 0 mm corresponds to a magic-momentum muon [4]. (b) The fitted distribution of vertical oscillation amplitudes before (dashed red line) and after (solid blue line) the azimuthal averaging and calorimeter acceptance corrections [4].

Driven by the vertical focusing provided by the ESQ system, the pitch correction is derived as $C_p \approx \frac{n}{2} \frac{\langle y^2 \rangle}{R_0^2} = \frac{n}{4} \frac{\langle A^2 \rangle}{R_0^2}$. The vertical oscillation amplitude $A$ is extracted from tracker measurements. Numerical simulations are used to model the azimuthal variations around the SR from the discrete ESQ sections. Moreover, not all decay positrons hit a calorimeter and enter into the determination of $\omega_a^m$. To account for this, each calorimeter acceptance has been modeled as a function of transverse and azimuthal decay position. Fig. 1b shows the fitted distribution of vertical oscillation amplitudes before and after these corrections. The resulting pitch correction is $C_p = 180 \pm 13$ ppb.

---

[1]Our coordinate system is with respect to the center of the storage volume at radius $R_0 = 7.112$ m, with $x$ or $r$ radially outward, $y$ vertically up, and $\phi$ increasing clockwise when viewed from above.

## 3. Muon Loss Correction $C_{ml}$ and Phase Acceptance Correction $C_{pa}$

The variable $\varphi_0$ in Eq. 2 depends on the stored muon momentum $p_\mu$, the decay positron energy $E$, and the transverse decay coordinates $(x, y)$ inside the SR. If the stored muon average transverse distribution and the detector gains are stable throughout a fill, that average phase remains constant. However, two resistors in the ESQ system were faulty, consequently causing slower-than-designed charging times during the first ~100 μs of the measurement period (see Fig. 2a). This led to a time-dependent phase shift from the correlations of decay position to average phase owing to the detector acceptance. An extensive study of this effect involved the determination of the time-dependent muon spatial distributions for each tracker station and for each dataset, and the evolution of these distributions using beam dynamics models and simulations to produce $M^c(x, y, t)$ for all azimuthal locations where calorimeters ($c$) are placed. To extract the time-dependent phase for each calorimeter $\varphi_{pa}^c(t)$, the $M^c(x, y, t)$ distributions are then combined with acceptance $\varepsilon^c$, asymmetry $A^c$, and phase maps $\varphi_{pa}^c$ using the following weighted sum over all spatial bins:

$$\varphi_{pa}^c(t) = \arctan \frac{\sum_{ij} M^c(x_i, y_j, t) \cdot \varepsilon^c(x_i, y_j) \cdot A^c(x_i, y_j) \cdot \sin(\varphi_{pa}^c(x_i, y_j))}{\sum_{ij} M^c(x_i, y_j, t) \cdot \varepsilon^c(x_i, y_j) \cdot A^c(x_i, y_j) \cdot \cos(\varphi_{pa}^c(x_i, y_j))}. \tag{4}$$

An example evaluation of $\varphi_{pa}^c(t)$ for the Run-1d dataset is shown in Fig. 2b. Simulated data are generated to evaluate the size of the mismeasurement of $\omega_a^m$ as a result of the time-dependent phase $\varphi_{pa}^c(t)$ by comparing the $\omega_a^m$ obtained with the fit function used to extract $\omega_a^m$. This difference yields the correction factor for each calorimeter. The phase-acceptance correction value is: $C_{pa} = 158 \pm 75$ ppb.

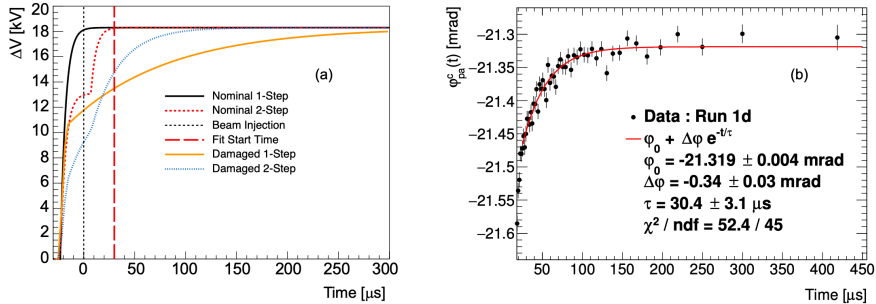

**Figure 2:** (a) Charging profiles for the nominal ESQ plates (black line and dotted red line). The two damaged resistors (solid orange and dotted blue lines) exhibit markedly different charging profiles during the data-fitting period [4]. (b) Calculation of $\varphi_{pa}^c(t)$ for Calorimeter 19 in Run-1d using Eq. 4 [4].

The muon loss correction takes into account the effect from muons that permanently escape from the storage volume during data taking and can potentially bias $\omega_a^m$ by inducing low drifts in the phase. A careful analysis demonstrated that lost muons do not significantly alter the ensemble-averaged spin phase, but we apply a small data-driven correction with $C_{ml} = -11 \pm 5$ ppb.

The damaged resistors were replaced after Run-1, which will significantly reduce the dominant contribution to $C_{pa}$ and the overall magnitude of muon losses in the next runs.

## 4. Conclusion

The key findings discussed in this proceeding are tabulated in Table 1. We described four beam dynamics systematic corrections that are required to adjust the measured muon precession frequency $\omega_a^m$ to its true physical value $\omega_a$. The pitch correction requires knowledge of the vertical stored muon distribution from the *in situ* tracker system. The electric field correction requires knowledge of the stored muon radial distribution, which is deduced from studying the time evolution of the incoming muon bunch. A small correction is applied for the muon loss-induced phase change. Finally, owing to two damaged high-voltage resistors in the ESQ system, the mean and rms of the stored muon distribution in Run-1 evolved throughout the first ∼100 μs of the measurement period. The sum of the corrections to $\omega_a^m$ is 0.50±0.09 ppm; the uncertainty is small compared to the 0.43 ppm statistical precision of $\omega_a^m$.

|                   | $C_e$ | $C_p$ | $C_{ml}$ | $C_{pa}$ | $C_{\text{total}}$ |
|-------------------|------|------|---------|---------|--------------------|
| Correction (ppb)  | 489  | 180  | −11     | −158    | 499                |
| Uncertainty (ppb) | 53   | 13   | 5       | 75      | 93                 |

**Table 1:** The Run-1 combined beam dynamics corrections to $\omega_a^m$ from the four Run-1 datasets.

## Acknowledgments

Work supported by Fermi Research Alliance, LLC under Contract No. DE-AC02-07CH11359 with the U.S. Department of Energy, Office of Science, Office of High Energy Physics.